\documentclass[12pt]{article}
\usepackage[T1]{fontenc}
\usepackage[english]{babel}
\usepackage[utf8]{inputenc}

\usepackage{rotating}
\usepackage{booktabs}
\usepackage{longtable}
\usepackage{natbib}
\usepackage{graphicx}
\usepackage{wrapfig}
\usepackage{setspace}
\usepackage{multirow}
\usepackage{bm}
\usepackage{amssymb,amsfonts,amsthm}

\usepackage[fleqn]{amsmath}

\usepackage{float}

\usepackage[section]{placeins}
\usepackage[titletoc,title]{appendix}
\usepackage{titletoc}
\usepackage[labelfont=bf]{caption}

\usepackage{hyperref}
\hypersetup{pdfborder={0 0 0}}

\usepackage[margin=1in]{geometry}
\usepackage{pdflscape}
\usepackage{enumitem}

\usepackage[hang,flushmargin]{footmisc}

\DeclareMathOperator{\E}{\mathbb{E}}

\DeclareMathOperator{\R}{\mathbb{R}}
\DeclareMathOperator{\p}{\mathrm{p}}

\usepackage{tikz}
\usetikzlibrary{calc}
\usetikzlibrary{matrix}
\usetikzlibrary{positioning}
\usetikzlibrary{arrows}
\usetikzlibrary{decorations.markings}
\usetikzlibrary{decorations.pathreplacing}
\usetikzlibrary{patterns}

\newtheorem{assumption}{Assumption}



\usepackage{siunitx}
\newcolumntype{d}{S[
    input-open-uncertainty=,
    input-close-uncertainty=,
    parse-numbers = false,
    table-align-text-pre=false,
    table-align-text-post=false
 ]}

 \usepackage{derivative}

\title{Bounding MrP: Subnational public opinion estimation without detailed census data}

\author{Moritz Marbach\thanks{\scriptsize University College London, \href{mailto:m.marbach@ucl.ac.uk}{m.marbach@ucl.ac.uk}. We are grateful for feedback on an earlier version of this paper from Lucas Leemann, Ben Lauderdale, and Alexander Coppock, as well as from participants in the panel ``Inferences in Space and Time'' at the 2026 EPSS conference. \newline \newline Large language models were used for proofreading, formatting, and drafting individual passages of the manuscript (May–July 2026), as well as for generating sections of the R code used in the empirical analysis (June 2025–May 2026). These tasks were performed using Anthropic's Claude Code, which automatically selected the underlying Anthropic language model. The specific model version was not recorded. All generated text and code was reviewed and revised by the author for accuracy, validity, and suitability before being incorporated into the manuscript and codebase.
}}

\begin{document}


\clearpage

\maketitle


\thispagestyle{empty}


\begin{abstract}
    Multilevel regression with post-stratification (MrP) is the dominant method for estimating public opinion in subnational units. The method requires the full joint distribution of demographic variables in a subnational unit, which is often unavailable from public census tabulations. Existing approaches fill this gap by imposing assumptions to combine marginal distributions of census variables into a synthetic joint distribution. We propose sharp MrP bounds instead: the smallest interval containing all estimates of subnational public opinion consistent with the marginal distributions alone. The bounds solve a pair of linear programs and are computationally feasible for typical applications in political science. Applied to two prominent examples in the literature, the bounds are tight enough to be informative and often more accurate than MrP restricted to a coarser, observed joint distribution of census variables. The bounds serve as margin-consistent interval estimates and a diagnostic when researchers use a synthetic joint distribution to construct MrP estimates.
\end{abstract}


\newpage
\section{Introduction}

Multilevel regression with post-stratification (MrP) has become the dominant method for estimating public opinion in small subnational units from survey data (e.g., US states, UK constituencies, Swiss cantons) where direct estimates are too noisy to be useful \citep{Gelman.Little.1997,Park.et.al.2004, Leemann.Wasserfallen.2020}. MrP proceeds in two steps. First, a multilevel model is fitted on survey data to predict public opinion as a function of demographic and contextual variables. Second, the fitted model predicts public opinion for each demographic stratum within a subnational unit, and these stratum-level predictions are then aggregated into a single unit-level estimate by weighting each stratum by its population share.

To construct the weights, a joint distribution of the post-stratification variables within each subnational unit is needed. Known as the post-stratification frame, this joint distribution provides the total number of persons for the cross-combination of demographic characteristics (e.g., the number of 25--34-year-old female university graduates in a particular county). While most countries have census data that could provide this joint distribution, privacy and disclosure rules typically limit small-area census tables to cross-tabulations of only a handful of variables. And even if the joint distribution is made available, most governments do not typically collect the variables most predictive of public opinion (e.g., party identification) as part of their censuses.

Aside from the often inadequate workaround of restricting the post-stratification step to the variables for which cross-tabulations from the census are available, more elaborate approaches have been proposed in the literature. One imputes missing variables into the post-stratification frame using an auxiliary model fitted on survey data. The model is constructed to predict the missing variable from variables available in the post-stratification frame---for instance, party identification is imputed using a model with variables for age, gender, and education \citep[e.g.,][]{Kastellec.et.al.2015}. Another approach combines multiple marginal distributions (e.g., age $\times$ gender and gender $\times$ education) into a ``synthetic joint distribution'' over the union of their variables (here, age $\times$ gender $\times$ education) by assuming statistical independence. An ``adjusted'' variant relaxes this independence assumption using the correlation structure between the demographic variables observed in a national survey \citep{Leemann.Wasserfallen.2017}. All these approaches construct a full distribution of post-stratification variables, but each does so by invoking assumptions about the unobserved joint distribution.

We present an alternative approach: rather than impose assumptions to construct the post-stratification frame from multiple marginal distributions, we treat the missing joint distribution as a partial identification problem \citep{Horowitz.Manski.1995, Manski.2003} and report the full range of MrP estimates that are consistent with the available cross-tabulations of post-stratification variables. The MrP estimand is a linear functional of the unknown post-stratification distribution, and the observed partial cross-tabulations impose linear equality constraints on that post-stratification distribution. The set of values the estimand can take is therefore an interval whose endpoints are obtained by minimising and maximising the empirical analogue of the estimand over all post-stratification distributions consistent with the observed margins. The resulting margin-consistent bounds of the MrP estimate (``MrP bounds'') are sharp: every interior value is attainable by some post-stratification distribution consistent with the observed tables, and no exterior value is attainable. The bounds reflect fundamental population-level identification uncertainty, which is distinct from the estimation uncertainty arising from the finite survey sample and measured in the form of confidence or Bayesian credible intervals.

Using our open-source R package\footnote{\url{https://github.com/sumtxt/mrpbounds}}, we illustrate the approach with the Swiss federal referendums analysed by \citet{Leemann.Wasserfallen.2017}, who construct a synthetic joint distribution from Swiss census data and electoral vote shares. For the subset of referendums they select for closer analysis, two findings stand out. First, the bounds are tight enough to discriminate subnational units with higher versus lower support for a proposed policy, without invoking additional assumptions about the unknown joint distribution of the post-stratification variables. Second, the achievable MSE range implied by the bounds often lies strictly below the MSE of a classic MrP estimate that relies on the Swiss census data alone, so incorporating the additional margins substantially improves predictions of referendum outcomes. Together, these results show that the bounds are valuable to researchers studying subnational public opinion or predicting subnational election outcomes from survey data, and that the assumptions needed to construct a synthetic joint distribution have only a limited influence on the resulting estimates in this application. We also discuss and illustrate the conditions under which these assumptions are likely to matter much more.

This research note makes two main contributions. First, we connect research on subnational public opinion estimation and multilevel modelling with post-stratification to the partial-identification framework of \citet{Horowitz.Manski.1995} and \citet{Manski.2003}, characterising the identified set as an interval computable with standard optimisation software. Different from many methodological innovations in the MrP literature in recent years \citep[e.g.,][]{Marble.Clinton.2026, Goplerud.Auslen.2025, Rosenman.et.al.2023, Broniecki.et.al.2022, Ornstein.2020}, we focus on the post-stratification step rather than improving the outcome model. Second, we show that these bounds serve a dual purpose: they provide a margin-consistent interval estimate of subnational public opinion, and they offer a diagnostic for the assumptions embedded in existing post-stratification approaches in the absence of detailed census data. Specifically, the bounds make legible how much violations of these assumptions could shift MrP estimates---a potential source of error that has so far been left opaque.

\section{Multilevel Regression with Post-stratification} \label{sec:mrp}

Multilevel regression with post-stratification (MrP) has emerged as the standard approach in political science for estimating public opinion in subnational units \citep{Lax.Phillips.2009,Warshaw.Rodden.2012}, but see \citet{Caughey.Warshaw.2019} for a more nuanced discussion.

Consider a national survey with respondents $i=1,\dots,n$ and an outcome $Y_i$. The researcher's goal is to estimate the population mean of $Y$ in some subnational unit $g$ (e.g.\ a state, region, or municipality),

\begin{equation}
\theta_g = \E[Y \mid G = g].
\end{equation}

MrP achieves this in two steps. First, a multilevel model is fitted on the survey to predict the outcome $Y_i$ as a function of post-stratification variables $X$.\footnote{Researchers often also include contextual variables describing the subnational unit; we omit this detail here, though the proposed approach accommodates it directly.} Next, the model is used to construct cell-level predictions $y_c = \E[Y \mid X = x_c, G = g]$ over post-stratification cells $c \in \mathcal{C}_g$ defined by the demographic variables $X$. Typically, $\mathcal{C}_g$ is constructed from census data. Each cell $c$ corresponds to a row of the joint cross-tabulation of $X$ for unit $g$, with $\pi_c$ the population share the census assigns to it. Second, cell means are aggregated using the within-unit joint distribution of $X$,

\begin{equation}
\label{eq:mrp-target}
\theta_g = \sum_{c\in \mathcal{C}_g} \pi_c \, y_c, \qquad \pi_c = \p(X = x_c \mid G=g).
\end{equation}

We refer to $\pi$ as the post-stratification distribution for unit~$g$.

Equation~\eqref{eq:mrp-target} makes clear that MrP requires the full joint distribution of the post-stratification variables within each unit. We are concerned with a setting in which the relevant census data reports only marginal tables (e.g.\ age $\times$ gender or education $\times$ region), but not the full cross-tabulation (e.g.\ age $\times$ gender $\times$ education $\times$ region).

\subsection{Existing Approaches: MrP Without Detailed Census Data} \label{sec:existing}

Much of the recent methodological progress in MrP focuses on the first step, the outcome model.\footnote{Examples of methodological innovations focused on the outcome model include hierarchical multilevel models with deep interactions \citep{Ghitza.Gelman.2013, Goplerud.2024}, multilevel models with contextual and spatial effects \citep{Selb.Munzert.2011, Buttice.Highton.2013, Hanretty.et.al.2018}, machine-learning models \citep{Ornstein.2020, Broniecki.et.al.2022}, extensions to multivariate outcomes \citep{Goplerud.Auslen.2025}, and post-estimation calibration of the predictions to known population-level quantities like vote shares \citep{Rosenman.et.al.2023, Marble.Clinton.2026}.} However, much less attention has been paid to the second step, post-stratification \citep[][p.9]{Caughey.et.al.2020}. With census-based post-stratification, the central problem is that the full joint distribution of all post-stratification variables (the post-stratification frame) is often unavailable \citep{Leemann.Wasserfallen.2017} as statistical agencies suppress detailed small-area cross-tabulations for privacy reasons. Moreover, censuses rarely collect the strongest predictors of public opinion, such as party identification.

The simplest workaround is to restrict MrP to variables for which a joint distribution is available. But omitting predictors from the multilevel model may lead to inconsistent MrP estimates \citep[see][for a discussion of post-stratification and auxiliary-variable choice]{Si.2025}. The most direct alternative is to construct a joint distribution from available population marginal distributions by assuming statistical independence. Under independence, cell counts are given by the product of marginal proportions and the unit population size. \citet{Leemann.Wasserfallen.2017} call the resulting joint distribution a ``simple synthetic joint distribution'' and the estimates ``simple MrsP'' (multilevel regression with \emph{synthetic} post-stratification). As \citet{Caughey.Wang.2019} emphasise, however, recovering a joint distribution from marginals is an ecological inference problem, and they stress that additional information from other data sources should be incorporated to avoid sensitivity to assumptions.

Iterative proportional fitting (IPF), or ``raking,'' is one way to incorporate this additional information \citep{Stephan.1942, Ireland.Kullback.1968, Little.Wu.1991}. Raking starts from a survey tabulation that records the full joint distribution of the post-stratification variables (often called the ``seed'') and adjusts this distribution iteratively so that its margins match the observed population margins. The resulting joint distribution then serves as the post-stratification frame. \citet{Leemann.Wasserfallen.2017} introduce a related approach that adjusts each margin only once rather than iterating to convergence. Because of this single pass, their ``adjusted synthetic joint distribution'' is not guaranteed to match all population margins. Since a post-stratification frame that matches the population margins is typically preferable, raking is the more attractive approach in practice.

Simple and adjusted synthetic joint distributions remain the two most popular approaches in political science.\footnote{Studies constructing a simple or adjusted synthetic joint distribution include \citet{Cerina.Duch.2025, Claassen.Gibson.2019, Hoover.2021, Zewail.et.al.2026, Franko.Pacheco.2025, Gravelle.2024, Lipps.Schraff.2021, Simonovits.Payson.2023, Huang.Liu.2025, Lauderdale.et.al.2020}.} A third approach, model-based imputation, instead models variables absent from the census in a national survey as functions of variables the census does record, and uses the fitted model to predict the distribution of the missing variable in the post-stratification frame \citep[see, for example][]{Kastellec.et.al.2015, Kuriwaki.et.al.2024}. To improve accuracy, one can constrain the model so that the implied marginal distribution of the imputed variable, after imputation into the frame, matches its observed population margin \citep{Kuriwaki.et.al.2024}.

When the joint distribution of post-stratification variables is unobserved, assumptions are necessary to combine the observed marginal distributions into a joint distribution. While raking and model-based imputation avoid the most stringent assumption (statistical independence), they still require two weaker ones. First, one must assume that the correlation structure substituted from survey data matches that in the population. Second, because subnational samples are too small to estimate the correlation structure locally, researchers tend to rely on a national sample and assume a common correlation structure across subnational units.

The consequences of these assumptions for MrP estimates are not well understood.\footnote{One exception is \citet{Ornstein.2020}, who shows in an appendix that when the outcome model is additively separable, simple MrsP yields the same post-stratified estimate as MrP with the unobserved true post-stratification, even if the independence assumption is violated. This result is of limited use for applied research, however, because applied work typically relies on outcome models that are not additively separable (e.g., a logistic regression with deep interactions).} The bounds we propose in the next section offer a substitute: instead of imposing additional assumptions to construct a point estimate, they report the range of estimates consistent with the observed margins alone. They also sidestep the question of how to propagate into final estimates the uncertainty introduced when using survey data to construct the post-stratification frame. In practice, political science research ignores this uncertainty, so that confidence (or credible) intervals are too short \citep[see also][on uncertainty propagation]{Li.Si.2024, Si.Zhou.2021}. That said, the bounds can also help researchers committed to point estimation make the impact of these assumptions on MrP estimates legible: large bounds mean that potential violations of the underlying assumptions are more consequential than narrow bounds are.

\section{Sharp Bounds on MrP Estimates} \label{sec:bounds}

Rather than making assumptions about the post-stratification distribution, we treat the missing joint distribution as a partial identification problem \citep{Horowitz.Manski.1995, Manski.2003}. We ask what range of values $\theta_g$ can take as $\pi$ varies over all distributions consistent with the observed marginal distributions. This range defines the sharp bounds for $\theta_g$: the smallest interval that contains all values of $\theta_g$ that are consistent with the observed data, without relying on assumptions beyond the observed margins.

Let $\mathcal{T} = \{T_1,\dots,T_J\}$ denote the set of observed marginal probability distributions for the post-stratification variables in unit $g$. Each $T_j$ imposes a set of linear equality constraints on $\pi$: the entries of $\pi$ summed over cells consistent with a given row of $T_j$ must equal the corresponding marginal probability. Together with the simplex conditions $\pi_c \geq 0$ and $\sum_c \pi_c = 1$, these constraints define the identified set of post-stratification distributions,

\begin{equation}
\label{eq:identified-set}
\mathcal{P}(\mathcal{T}) \;=\; \big\{ \pi \in \R^{|\mathcal{C}_g|} : \; A\pi = b, \;\; \pi \geq 0 \big\},
\end{equation}

where the constraint matrix $A$ and the right-hand side $b$ collect the simplex and tabular constraints implied by $\mathcal{T}$ (see Appendix~\ref{sec:app-constraints} for details).

Because $y_c$ does not depend on $\pi$, the MrP estimand in~\eqref{eq:mrp-target} is a linear functional of $\pi$. The range of values it can take as $\pi$ varies over $\mathcal{P}(\mathcal{T})$ equals:

\begin{equation}
\label{eq:lp}
\underline{\theta}_g \;=\; \min_{\pi\in\mathcal{P}(\mathcal{T})} \sum_{c} \pi_c y_c, \qquad
\overline{\theta}_g \;=\; \max_{\pi\in\mathcal{P}(\mathcal{T})} \sum_{c} \pi_c y_c.
\end{equation}

The interval $[\underline{\theta}_g, \overline{\theta}_g]$ is the identified set for the MrP estimand. We refer to this interval as the MrP bounds.

As we show in Appendix \ref{sec:app-sharpness}, the interval is sharp, i.e., every value inside is attained by some $\pi$ consistent with the observed margins, and no value outside is attained. It is also the smallest possible interval that does not rely on assumptions beyond those imposed by the observed marginal distributions. The same logic underlies the classical Fr\'echet--Hoeffding bounds on a joint distribution with fixed marginals \citep{Frechet.1951, Hoeffding.1940, Ruschendorf1991}.

By construction, any MrP estimate constructed from a synthetic post-stratification frame will lie within the bounds whenever that frame's margins are consistent with the margins used to construct the bounds. In practice, this means that MrP estimates will always lie within the MrP bounds, whether based on a simple synthetic post-stratification frame built under the independence assumption or on an adjusted frame built by raking survey data toward those margins.

There are two straightforward extensions of these bounds. First, the bounds generalise to any other quantity that can be written as a linear functional of $\pi$, e.g.\ differences between two MrP estimates ($\theta_g - \theta_g'$), or some rescaled population proportions ($\theta_g/r$ with $r$ as a constant). Second, any additional side information that can be written as a linear restriction on $\pi$ enters the constraint system on equal footing with the observed tables. Cells known a priori to be empty in the population (e.g., structural zeros such as married minors or under-10s holding doctoral degrees) can be encoded as equality constraints $\pi_c = 0$, and partial knowledge of the form $\pi_c \leq \bar{\pi}_c$ or $\pi_c \geq \underline{\pi}_c$ as additional inequalities. This contrasts with existing approaches, which by construction assign positive probability to every cell compatible with the margins and may therefore place mass on cells that cannot occur in the population.

\subsection{Estimation and Statistical Inference}

Estimation and inference can proceed along frequentist or Bayesian lines. Because most MrP applications in political science rely on simulation-based Bayesian inference, we focus on the latter. With simulation-based Bayesian inference, the user specifies a model for the survey data (the likelihood), assigns beliefs for all parameters (the prior), and then draws samples from the resulting posterior distribution of the parameters. The posterior draws for the parameters are used to construct a posterior distribution for the relevant cell predictions. We write those as $\{y_c^{(s)}\}_{s=1}^{S}$. If the post-stratification distribution is known, the draws from the posterior distribution of $y_c$ are combined with the post-stratification frame to obtain the posterior distribution of $\theta_g$. The mean of the posterior distribution is typically used as the MrP estimate, and the $95\%$ central credible interval is used to characterise the estimation uncertainty.

When the post-stratification distribution is unknown, nothing changes when it comes to the multilevel model. But instead of combining the posterior distribution with a known post-stratification distribution, one solves the two linear programs in equation~\eqref{eq:lp} once per draw, replacing $y_c$ by $y_c^{(s)}$ and leaving the constraint system unchanged. Both programs are linear (i.e., the objective is linear in $\pi$, and the tabular, simplex, and non-negativity restrictions are linear equalities and inequalities), and $\mathcal{P}(\mathcal{T})$ is a compact polytope, so an optimum is always attained at a vertex and off-the-shelf solvers return global optima. This yields a posterior distribution of the interval $[\underline{\theta}_g, \overline{\theta}_g]$.

To characterise the posterior distribution of the interval, we propose to report two quantities. The first is the interval running from the posterior mean of the lower bound to the posterior mean of the upper bound (\emph{set of posterior means}). This is the set-valued analogue of the posterior mean and serves as an estimate for the MrP bounds. As a second quantity, we propose to report the shortest interval that entirely contains $95\%$ of the posterior draws of the interval $[\underline{\theta}^{(s)}_g, \overline{\theta}^{(s)}_g]$. This is a \emph{credible set} for the identified set and covers it with posterior probability $0.95$. It is the set-valued analogue of the credible interval.

In addition, it is possible to make probability statements over MrP bounds. One can compute the posterior probability that every value within the bounds falls below some constant $c$, that is, $\p([\underline{\theta}_g, \overline{\theta}_g] \subseteq (-\infty, c))$, and the posterior probability that some value does, that is, $\p([\underline{\theta}_g, \overline{\theta}_g] \cap (-\infty, c) \neq \emptyset)$. The first says how confident one can be that, for example, support for a policy falls short of $c$ whatever the post-stratification distribution turns out to be; the second says whether the data still leave that possible.  The same logic applies symmetrically to $(c, \infty)$, allowing analogous statements about how confident one can be that support for a policy exceeds the threshold $c$.

The approach taken here can be viewed as an instance of the framework of \citet{Kline.Tamer.2016}, with the vector of cell predictions $\{y_c\}_{c \in \mathcal{C}_g}$ playing the role of their point-identified reduced-form parameter and the two linear programs in~\eqref{eq:lp} playing the role of their identified-set mapping. We share their focus on the identified set rather than on the partially identified parameter $\theta_g$. We note that \citeauthor{Kline.Tamer.2016} also provide conditions under which the proposed credible set is an exact frequentist confidence set for the identified set, in the sense that it covers the identified set with the nominal probability asymptotically. We point to \citet{Giacomini.Kitagawa.2021} for a proof that the proposed set of posterior means converges to the identified set asymptotically.

\section{Application} \label{sec:illustration}

Next, we illustrate the proposed approach with Swiss federal referendums analysed by \citet{Leemann.Wasserfallen.2017}, who construct public opinion estimates across 186 referendums and 26 Swiss cantons; we discuss a second example using US data in Appendix~\ref{sec:app-us}. Their multilevel probit model includes individual-level random effects for gender, age group, education, and party identification, alongside random effects for cantons and regions, and two cantonal contextual variables (German-language share and Catholic share). The authors fit the model separately for each referendum using maximum likelihood. Because the census does not record party identification, they construct a synthetic joint distribution by combining the marginal distribution of gender, age, education, and canton from the Swiss census with the marginal distribution of cantonal party vote shares.

\begin{figure}[htbp]
    \centering
    \caption{MrP Bounds and Estimates for 4 Swiss Referendums}
    \label{fig1}
    \includegraphics[width=\textwidth]{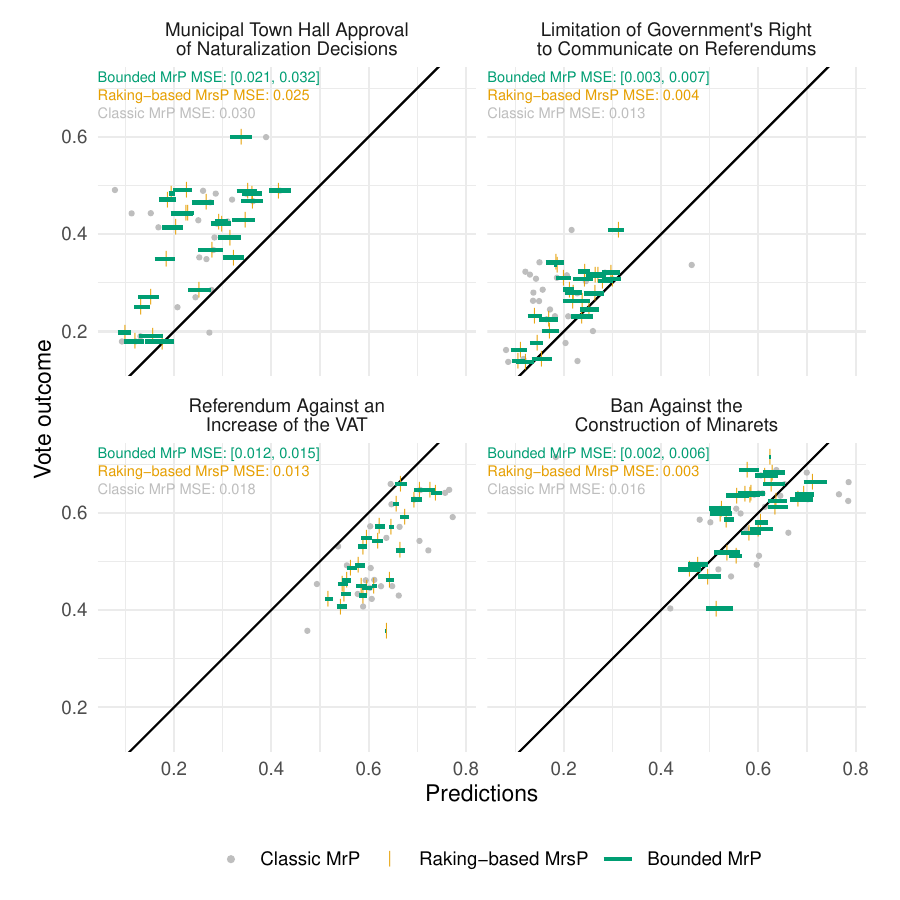}
    \endgraf\vspace{1em}
    \begin{minipage}{\linewidth} \scriptsize
    \textbf{Notes}: Each scatter plot panel corresponds to one referendum and plots posterior-mean MrP/MrsP estimates against the observed cantonal yes-vote share for each of the 26 Swiss cantons. The gray dot marks the classic MrP estimate, the orange tick marks the raking-based MrsP estimate, and the horizontal green bar spans the MrP bounds. The black diagonal marks perfect prediction. The top-left corner reports, from top to bottom, the achievable MSE range for bounded MrP, the MSE of the raking-based MrsP estimates, and the MSE of the classic MrP estimates.
    \end{minipage}
\end{figure}

They benchmark their MrsP estimates against classic MrP estimates, which they construct by post-stratifying using only the known joint distribution of census variables (age, gender, and education) and including the vote shares as contextual variables. To assess performance, Leemann and Wasserfallen compare both sets of estimates against the yes-vote shares. They focus their main analysis on four selected cantonal referendums: municipal-level approval of naturalisation decisions, limits on government communication during referendum campaigns, a VAT increase, and a ban on new minaret construction. Across these four referendums, the authors find that the mean-squared error (MSE) between their MrsP estimates and the yes-vote shares is lower than the MSE of the classic MrP estimates that rely on the joint distribution of census variables alone.

To illustrate the proposed MrP bounds, we reestimate the authors' model specification on the same data using simulation-based Bayesian inference. Appendix~\ref{sec:app-swiss-bayes} details our multilevel logistic model, priors, and convergence diagnostics. For each canton and referendum, we solve the two linear programs in equation \eqref{eq:lp} once per posterior draw, using two tables: the joint distribution of the census variables (age, gender, and education) and the cantonal party vote shares from the federal election. As proposed in the previous section, we estimate the MrP bounds using the set of posterior means, and we compare it to the classic MrP estimate and the raking-based MrsP estimate, both likewise computed as posterior means from the same draws. In Appendix~\ref{sec:lw17comp}, we compare our raking-based MrsP estimates with the MrsP estimates reported by \citet{Leemann.Wasserfallen.2017}. As expected, the two are highly correlated, but we prefer the raking-based MrsP estimates because they match the observed population margins exactly, as discussed in Section \ref{sec:existing}.

Figure~\ref{fig1} displays the results. Each panel corresponds to one referendum. The horizontal thick green bar spans the MrP bounds of each canton. In addition, the figure shows two point estimates: the vertical orange bar marks the raking-based MrsP estimate, while the gray dot marks the classic MrP estimate. Following the approach of \citet{Leemann.Wasserfallen.2017}, we report three MSE statistics in the top-left corner of each panel: from top to bottom, the achievable MSE range for bounded MrP\footnote{The MSE bounds give the minimum and maximum MSE achievable by choosing, for each canton, any estimate within the bounds $[\underline{\theta}_g, \overline{\theta}_g]$. Since squared error is convex, each canton's contribution can be optimised independently: the minimum is zero whenever the official yes-vote share $v_g$ falls within the interval, and the squared distance to the nearer endpoint otherwise; the maximum is always the squared distance to the farther endpoint. The MSE bounds are the averages of these per-canton optima across the 26 cantons.}, the MSE of the raking-based MrsP estimates, as well as the MSE of the classic MrP estimates.

Two findings stand out. First, the raking-based MrsP point estimates fall within the MrP bounds, and their MSE falls within the achievable MSE range. More importantly, the achievable MSE range for bounded MrP lies strictly below the MSE of the classic MrP estimates in three out of four referendums. So while the marginal distributions alone do not yield a point estimate of public opinion in a canton, they produce a substantially more informative interval estimate than a classic MrP point estimate. Second, inspecting the scatterplot, we observe that the widths of the bounds are modest relative to the cross-cantonal variation in support. The observed margins tightly constrain the set of consistent MrP estimates under the model. This is good news: applied researchers interested in characterising subnational public opinion can do so with the bounds without committing to the assumptions needed to reconstruct a synthetic post-stratification frame from observed margins.

That said, Figure~\ref{fig1} omits the estimation uncertainty. In Appendix Figure \ref{fig:ppint}, we report the MrP bounds alongside their 95\% credible set and the MrsP estimates alongside their 95\% credible intervals. We observe that the width of the credible set is only slightly larger than the credible interval. Different from the credible interval, which only measures estimation uncertainty, the credible set is wider as it also includes the identification uncertainty introduced by the absence of the joint distribution of the post-stratification frame.

The narrow MrP bounds observed here will not always arise: their width depends on three factors: how probability mass is distributed across the known margins of the post-stratification distribution, the correlations between the post-stratification variables and the outcome, and the structure of the multilevel model. One reason the bounds are narrow here is that the authors' model is additive in the post-stratification variables, with no interactions between party identification and the other census variables. As we confirm in Appendix~\ref{sec:app-stress}, adding such interactions widens the bounds substantially, since there is no information on the dependency between party identification and the other census variables.

In Appendix~\ref{sec:app-us}, we report a secondary application to US same-sex marriage support across 174 state senate districts \citep{Warshaw.Rodden.2012}. As in \citet{Leemann.Wasserfallen.2017} and different from the analysis above, adding age as a post-stratification variable does not meaningfully improve on classic MrP, because age is far less predictive of attitudes than party identification in the analysis above. Correspondingly, the MrP bounds overlap strongly with classic MrP estimates.

\section{Discussion and Conclusion} \label{sec:conclusion}

This paper takes a different route to a familiar problem. When the full joint distribution of demographic variables is unavailable, applications of MrP have typically restricted themselves to the subset of post-stratification variables for which a joint distribution is available or reconstructed the joint distribution under additional assumptions. We propose instead to treat the missing joint distribution as a partial identification problem, reporting the full range of MrP estimates consistent with the marginal distributions via sharp, computationally feasible linear-programming bounds. We illustrate the MrP bounds with data on Swiss referendums \citep{Leemann.Wasserfallen.2017} and US same-sex marriage \citep{Warshaw.Rodden.2012}. We confirm that the bounds are tight and informative. Beyond serving as interval estimates in their own right, the bounds double as a diagnostic for researchers who prefer point estimates: the larger the bounds, the stronger the assumptions invoked to pin down a raking-based MrsP estimate relative to the range of feasible values.

Our approach has limitations. The number of post-stratification cells grows multiplicatively in variables and their levels, so very large frames may eventually strain open-source solvers. However, existing MrP applications typically rely on few enough post-stratification variables that this constraint rarely binds. The bounds also do not discriminate between distributions inside the identified set: researchers who require a single number must still impose additional assumptions, and the bounds are silent on which are most defensible. The midpoint is not a natural default. Unlike a confidence interval, the identified set has no natural centre. In some cases the lower or upper bound alone may be defensible, as it describes a ``worst-case scenario'' for a given policy question.

This paper adapts the partial identification framework of \citet{Horowitz.Manski.1995} and \citet{Manski.2003} to the MrP setting. Partial identification remains a niche area of research in political science \citep[though see][]{Duarte.et.al.2024, Knox.et.al.2020}, perhaps because such bounds are perceived as uninformatively wide. As we demonstrate here, however, the MrP bounds in both applications are narrow and therefore informative about subnational public opinion. More generally, the bounds express uncertainty from limited data availability, just as confidence (or credible) intervals express sampling uncertainty. As a field, we should become more comfortable with both types of uncertainty.

\newpage

\addcontentsline{toc}{section}{References}
\begin{spacing}{2}
\bibliographystyle{apsr}
\bibliography{ref}
\end{spacing}

\newpage

\begin{appendices}

\renewcommand\thetable{SM.\arabic{table}}
\renewcommand\thefigure{SM.\arabic{figure}}
\renewcommand\theequation{SM.\arabic{equation}}

\setcounter{figure}{0}
\setcounter{table}{0}
\setcounter{equation}{0}

\begin{center}
{\LARGE\bfseries Supplementary Information: Bounding MrP: Subnational public opinion estimation without detailed census data}
\end{center}

\bigskip

\startcontents[appendixtoc]
\printcontents[appendixtoc]{}{1}{}

\newpage

\section{Additional Formal Details}

\subsection{Algorithmic Construction of the Constraint System} \label{sec:app-constraints}

To construct the constraint system $A\pi = b$ that appears in the linear program, one proceeds in two steps. First, enumerate the cells of the post-stratification table as the Cartesian product of all variable levels observed across the input tables. With $D = \prod_{k=1}^K L_k$ cells in total, where $L_k$ is the number of levels of variable $k$, the unknown distribution $\pi$ is a vector in $\mathbb{R}^D$.

Second, build one row of $A$ for every row of every input table, plus one extra row that requires $\pi$ to sum to one. A constraint row is the indicator of the cells that agree with the corresponding table row: it places ones at every cell whose values match the table row on the variables described by the table, and zeros elsewhere. The right-hand-side entry of $b$ is the probability the table records for that row. The last row of each table is dropped, since it is implied by the sum-to-one row and the preceding rows.

\paragraph{Example.} Suppose there are three binary variables $V_1, V_2, V_3$, so $D = 8$ cells. Let $T_{12}$ denote the observed joint table for $(V_1, V_2)$, with entries $T_{12}(v_1, v_2) = \p(V_1 = v_1, V_2 = v_2)$, and let $T_3$ denote the observed marginal table for $V_3$, with entries $T_3(v_3) = \p(V_3 = v_3)$. After dropping the last row of each, $T_{12}$ contributes three constraints and $T_3$ contributes one; together with the sum-to-one row, the system has five rows:

\begin{equation}
\label{eq:example-constraint}
A \;=\;
\begin{pmatrix}
1 & 1 & 1 & 1 & 1 & 1 & 1 & 1 \\
1 & 1 & 0 & 0 & 0 & 0 & 0 & 0 \\
0 & 0 & 1 & 1 & 0 & 0 & 0 & 0 \\
0 & 0 & 0 & 0 & 1 & 1 & 0 & 0 \\
1 & 0 & 1 & 0 & 1 & 0 & 1 & 0
\end{pmatrix},
\qquad
b \;=\;
\begin{pmatrix}
1 \\
T_{12}(1,1) \\
T_{12}(1,2) \\
T_{12}(2,1) \\
T_{3}(1)
\end{pmatrix},
\end{equation}

where cells are ordered $(V_1,V_2,V_3) = (1,1,1), (1,1,2), (1,2,1), \dots, (2,2,2)$. Figure~\ref{fig:joint-table-correspondence} makes the correspondence concrete. The two $2 \times 2$ slices on top arrange the eight cells $\pi_1, \dots, \pi_8$ by the level of $V_3$; the equations below restate each row of $A\pi = b$ as a sum over cells.

\begin{figure}[ht]
    \centering
    \caption{Correspondence Between the Joint Table and the Constraint System}
    \label{fig:joint-table-correspondence}
    \setlength{\tabcolsep}{8pt}
    \renewcommand{\arraystretch}{1.15}
    \fbox{%
    \begin{minipage}{0.95\linewidth}
    \centering
    \vspace{0.75em}

    \begin{tabular}{cc}
    \textbf{$V_3 = 1$} & \textbf{$V_3 = 2$} \\[2pt]
    \begin{tabular}{c|cc}
                    & $V_2{=}1$ & $V_2{=}2$ \\ \hline
    $V_1{=}1$       & $\pi_1$   & $\pi_3$   \\
    $V_1{=}2$       & $\pi_5$   & $\pi_7$   \\
    \end{tabular}
    &
    \begin{tabular}{c|cc}
                    & $V_2{=}1$ & $V_2{=}2$ \\ \hline
    $V_1{=}1$       & $\pi_2$   & $\pi_4$   \\
    $V_1{=}2$       & $\pi_6$   & $\pi_8$   \\
    \end{tabular}
    \end{tabular}

    \vspace{1em}

    \begin{tabular}{r@{$\;=\;$}l@{\hspace{1.5em}}l}
    $\pi_1 + \pi_2 + \pi_3 + \pi_4 + \pi_5 + \pi_6 + \pi_7 + \pi_8$ & $1$          \\
    $\pi_1 + \pi_2$                                                & $T_{12}(1,1)$ \\
    $\pi_3 + \pi_4$                                                & $T_{12}(1,2)$ \\
    $\pi_5 + \pi_6$                                                & $T_{12}(2,1)$ \\
    $\pi_1 + \pi_3 + \pi_5 + \pi_7$                                & $T_{3}(1)$    \\
    \end{tabular}

    \vspace{0.75em}
    \end{minipage}%
    }
    \endgraf\vspace{1em}
    \begin{minipage}{0.8\linewidth}\scriptsize
    \textbf{Notes:} The two slices on top show the eight cells of the joint distribution of $(V_1, V_2, V_3)$. Each equation below states one row of $A \pi = b$: the cells that appear on the left-hand side are exactly those for which the corresponding row of $A$ has a one.
    \end{minipage}
\end{figure}

\subsection{Sharpness of the bounds} \label{sec:app-sharpness}

\begin{assumption}
\label{assn:feasible}
The observed tables in $\mathcal{T}$ are mutually compatible: there exists at least one probability distribution over $\mathcal{C}_g$ whose margins agree with every $T_j \in \mathcal{T}$.
\end{assumption}

Equivalently, the identified set $\mathcal{P}(\mathcal{T})$ is non-empty. Compatibility is automatic whenever the tables describe disjoint sets of variables, and is satisfied by construction when every $T_j$ is derived from the same underlying data.

To demonstrate that the interval $[\underline{\theta}_g, \overline{\theta}_g]$ defined in~\eqref{eq:lp} is \emph{sharp} (i.e., every value inside the interval is attained by some post-stratification distribution $\pi$ consistent with the observed tables, and no value outside is), we use standard convex analysis \citep[see][Ch.~2--4]{Boyd.Vandenberghe.2004}. Under Assumption~\ref{assn:feasible}, $\mathcal{P}(\mathcal{T})$ is a non-empty, compact, convex polytope contained in the simplex. The functional $\pi \mapsto \sum_c \pi_c y_c$ is linear, so its image on $\mathcal{P}(\mathcal{T})$ is a compact convex subset of $\mathbb{R}$, i.e. a closed interval. The endpoints of this interval are the linear-programming optima $\underline{\theta}_g$ and $\overline{\theta}_g$ defined in equation~\eqref{eq:lp}. Therefore every $\theta \in [\underline{\theta}_g, \overline{\theta}_g]$ is realised by some $\pi \in \mathcal{P}(\mathcal{T})$, and no value outside is consistent with the constraints, i.e. the interval is sharp.

\clearpage \section{Additional Empirical Results}

\subsection{Estimation and Posterior Uncertainty} \label{sec:app-swiss-bayes}

We estimate the model via Hamiltonian Monte Carlo (HMC) using \citet{Gabry.Goodrich.2016} rather than maximum likelihood. We use the default weakly informative priors on the model parameters and run 4 chains in parallel. We discard the first 1{,}000 draws per chain as burn-in and keep the remaining 1{,}000 per chain. Standard convergence diagnostics produce no evidence of non-convergence in the HMC chains.

We use a sample of 1{,}000 posterior draws per referendum and apply the linear programs in~\eqref{eq:lp} to each draw separately, yielding 1{,}000 lower--upper bound pairs per canton and referendum. The constraint system with census cross-tabulation (age $\times$ gender $\times$ education $\times$ canton) and cantonal party vote shares is identical to the frequentist analysis; only the objective vector changes across draws.

\begin{figure}[ht]
    \centering
    \caption{Posterior Distributions of MrP Bounds for 4 Swiss Referendums}
    \label{fig:ppdist}
    \includegraphics[width=\textwidth]{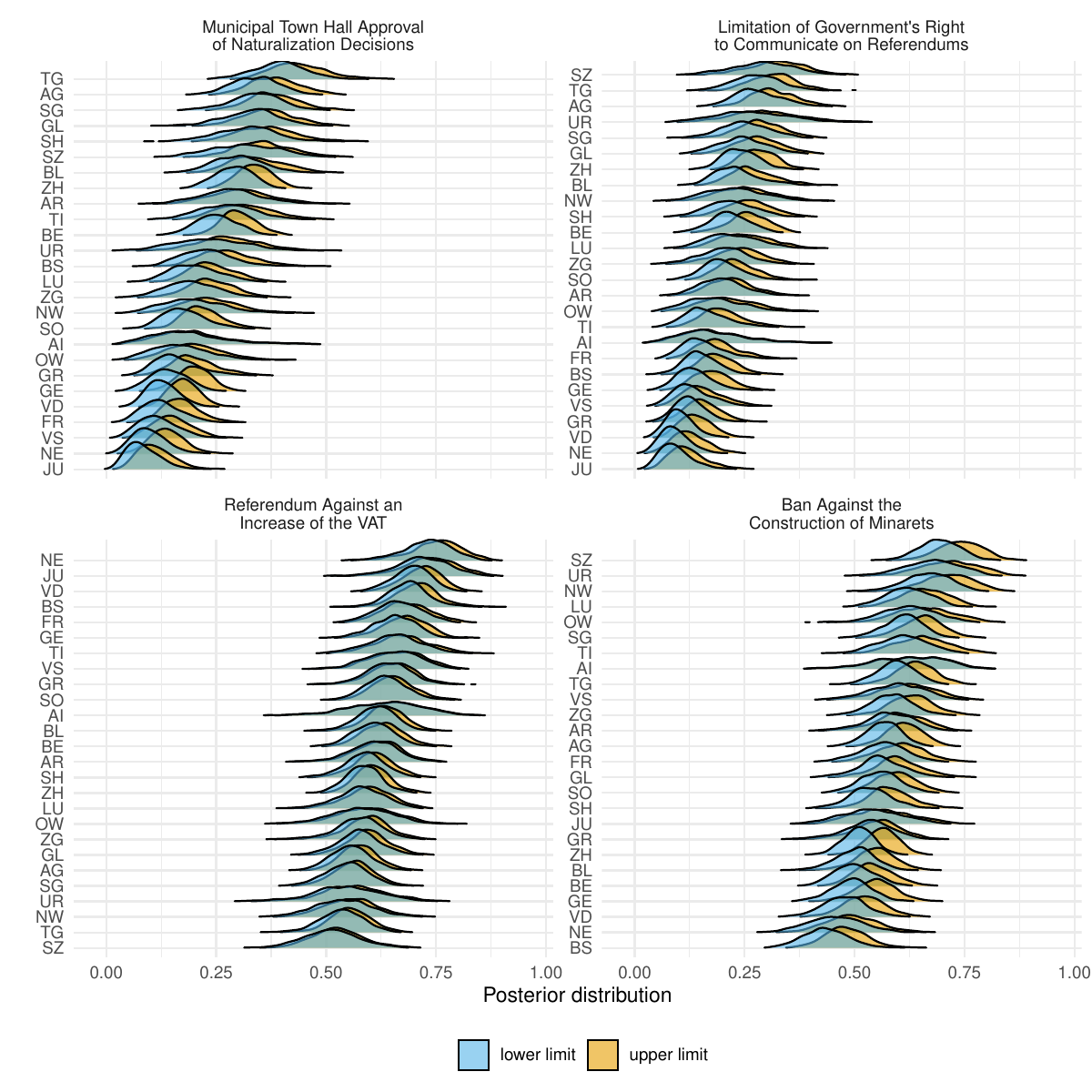}
    \endgraf\vspace{1em}
    \begin{minipage}{\linewidth}\scriptsize
    \textbf{Notes:} Each panel corresponds to one referendum and shows, for each of the 26 Swiss cantons, the blue density for the posterior of the lower MrP bound and the yellow density for the posterior of the upper bound. Cantons are ordered within each panel by the posterior median of the lower bound.
    \end{minipage}
\end{figure}

\begin{figure}[ht]
    \centering
    \caption{MrP Bounds and Raking-Based MrsP Estimates with 95\% Credible Sets and Credible Intervals for 4 Swiss Referendums}
    \label{fig:ppint}
    \includegraphics[width=\textwidth]{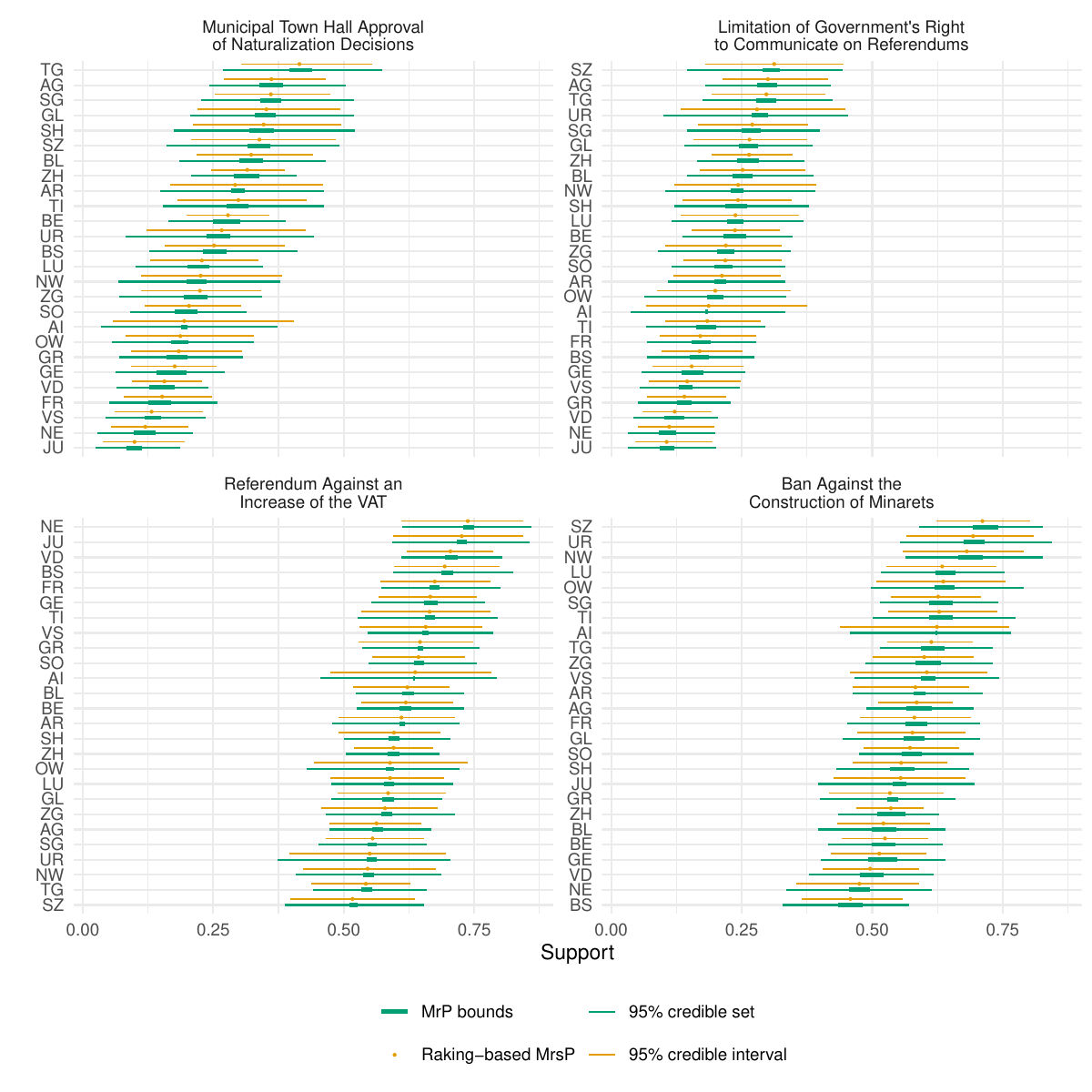}
    \endgraf\vspace{1em}
    \begin{minipage}{\linewidth}\scriptsize
    \textbf{Notes:} Each panel corresponds to one referendum and shows, for each of the 26 Swiss cantons, the posterior uncertainty of the MrP bounds and the raking-based MrsP estimate. The thick green bar marks the MrP bounds (the set of posterior means of the lower and upper bound), and the thin green line marks the 95\% credible set (the shortest interval containing 95\% of the posterior draws of the bounds). The orange dot marks the posterior mean of the raking-based MrsP estimate, and the thin orange line marks its 95\% credible interval. Cantons are ordered within each panel by the posterior mean of the raking-based MrsP estimate.
    \end{minipage}
\end{figure}

Figure~\ref{fig:ppdist} displays the resulting posterior distributions. For each canton and referendum, the blue density shows the posterior of the lower bound and the yellow density shows the posterior of the upper bound. Figure~\ref{fig:ppint} displays the MrP bounds and the raking-based MrsP estimate for each canton, together with the $95\%$ credible set for the bounds and the $95\%$ credible interval for the MrsP estimate.

\subsection{Comparing Adjusted MrsP with Raking-based MrsP} \label{sec:lw17comp}

\begin{figure}[ht]
    \centering
    \caption{MrsP for 4 Swiss Referendums}
    \label{fig:rep}
    \includegraphics[width=\textwidth]{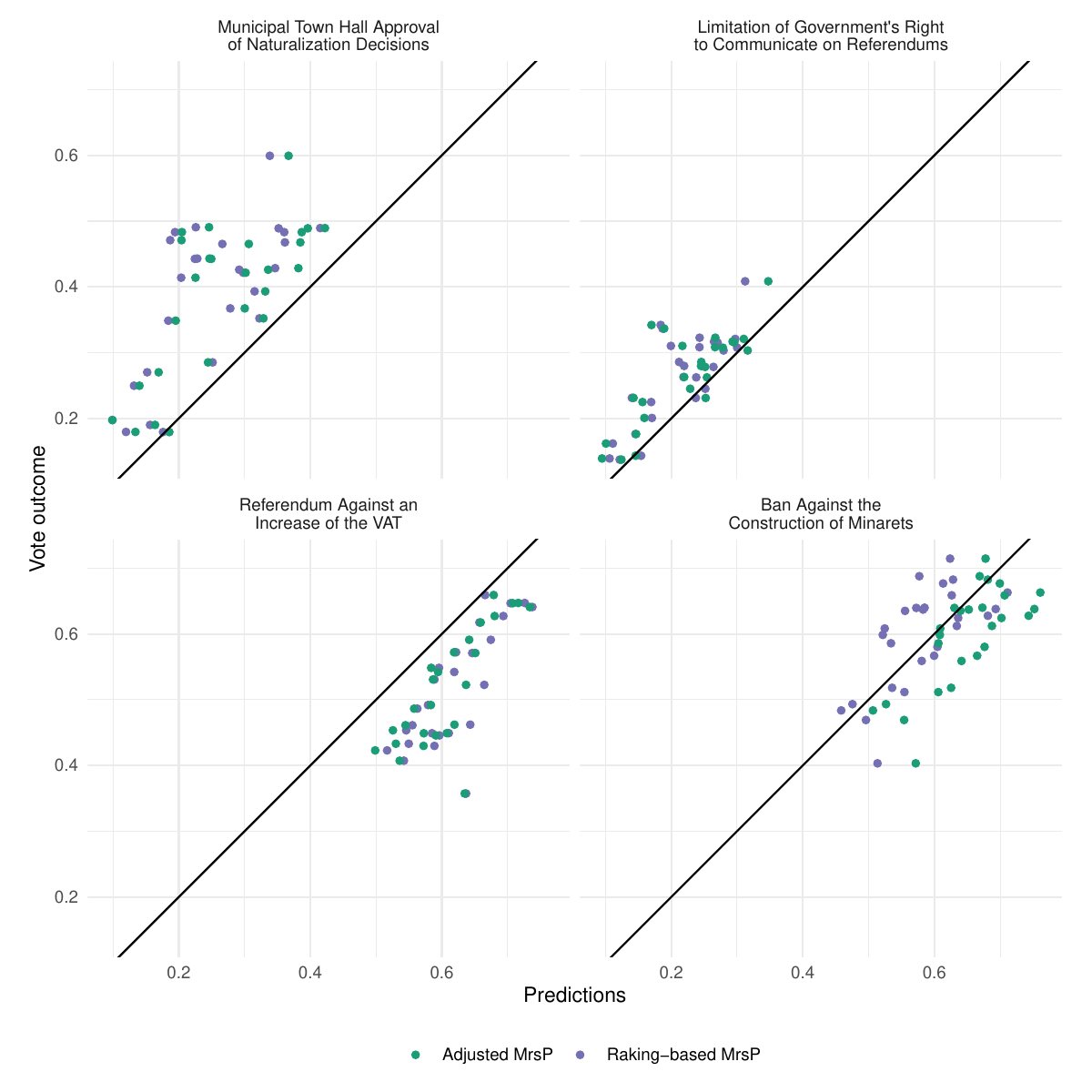}
    \endgraf\vspace{1em}
    \begin{minipage}{\linewidth}\scriptsize
    \textbf{Notes:} Each scatter plot panel corresponds to one referendum and plots MrsP estimates against the observed cantonal yes-vote share. Each point marks one of the 26 Swiss cantons: green points show our raking-based MrsP estimate and purple points show the adjusted MrsP estimate from \citet{Leemann.Wasserfallen.2017}. The black diagonal marks perfect prediction.
    \end{minipage}
\end{figure}

\citet{Leemann.Wasserfallen.2017} present adjusted MrsP estimates that diverge from the raking-based MrsP estimates we present in the main text. Figure \ref{fig:rep} shows the scatterplot of their adjusted MrsP and our raking-based MrsP estimates. The raking-based MrsP estimates rely on a synthetic joint distribution constructed with iterative proportional fitting (IPF), which rescales the cross-tabulation of survey data to match one target margin at a time, cycling through all margins until each is matched simultaneously, i.e., until convergence.

Deviations from the raking-based MrsP are expected because the Leemann and Wasserfallen adjusted MrsP applies only the first iteration of IPF, which does not guarantee that the margins of the synthetic joint distribution match the observed population margins. Moreover, a row normalisation is missing in their replication code, which further amplifies the difference.\footnote{The coding bug is located in file \texttt{source\_MrsP\_CH\_adjustedsynthetic.R}. In line 171, \texttt{party.ideal.corr} should be normalised, i.e., \texttt{party.ideal.corr <- party.ideal.corr / rowSums(party.ideal.corr)} as it is done in the corresponding script \texttt{aaa MrP and MrsP\_US.R} where it reads: \texttt{Age.district.percent <- Age.district/rowSums(Age.district)}.} Across the four referendums, the two MrsP estimates are highly correlated with each other (Pearson $r=0.99$) and with the referendum outcomes: the raking-based MrsP correlates with the vote outcome at $r=0.86$, compared to $r=0.89$ for the Leemann and Wasserfallen estimates. This modest gap is largely driven by the referendum on banning the construction of minarets (see the bottom-right panel in Figure \ref{fig:rep}).

\subsection{Expanding the Model} \label{sec:app-stress}

The multilevel model used in the main text is additive in all variables: party identification, gender, education, age, canton, and region, with no interaction between party identification and the other post-stratification variables. To test how much this additivity drives the narrow bounds reported in Figure~\ref{fig1}, we re-estimate the model after adding three interaction terms between party identification and each census-based post-stratification variable (gender, education, and age), while leaving the remaining specification and estimation unchanged. These interactions let the effect of party identification vary by gender, education, and age. As in the baseline model, we model the distribution of each interaction term with a random effect. Formally, the model takes the following form:

\begin{align}
\Pr(Y_i = 1) = \text{logit}^{-1}\Big(&\beta_0 + \beta_1 \, \text{German}_{\text{canton}[i]} + \beta_2 \, \text{Catholic}_{\text{canton}[i]} \notag \\
&+ \alpha^{\text{gender}}_{\text{gender}[i]} + \alpha^{\text{edu}}_{\text{edu}[i]} + \alpha^{\text{age}}_{\text{age}[i]} + \alpha^{\text{party}}_{\text{party}[i]} + \alpha^{\text{canton}}_{\text{canton}[i]} + \alpha^{\text{region}}_{\text{region}[i]} \notag \\
&+ \underbrace{\alpha^{\text{gender:party}}_{\text{gender}[i],\text{party}[i]} + \alpha^{\text{edu:party}}_{\text{edu}[i],\text{party}[i]} + \alpha^{\text{age:party}}_{\text{age}[i],\text{party}[i]}}_{\text{expanded model only}}\Big),
\end{align}
where each set of effects is drawn from its own group-level normal distribution, $\alpha^{k}_{j} \sim N(0,\sigma_k^2)$, independently across levels $j$ of grouping factor $k$. The baseline model omits the three interaction terms; the expanded model includes them.

\begin{figure}[ht]
    \centering
    \caption{MrP Bounds Under an Expanded Model with Interactions}
    \label{fig:app-stress}
    \includegraphics[width=\textwidth]{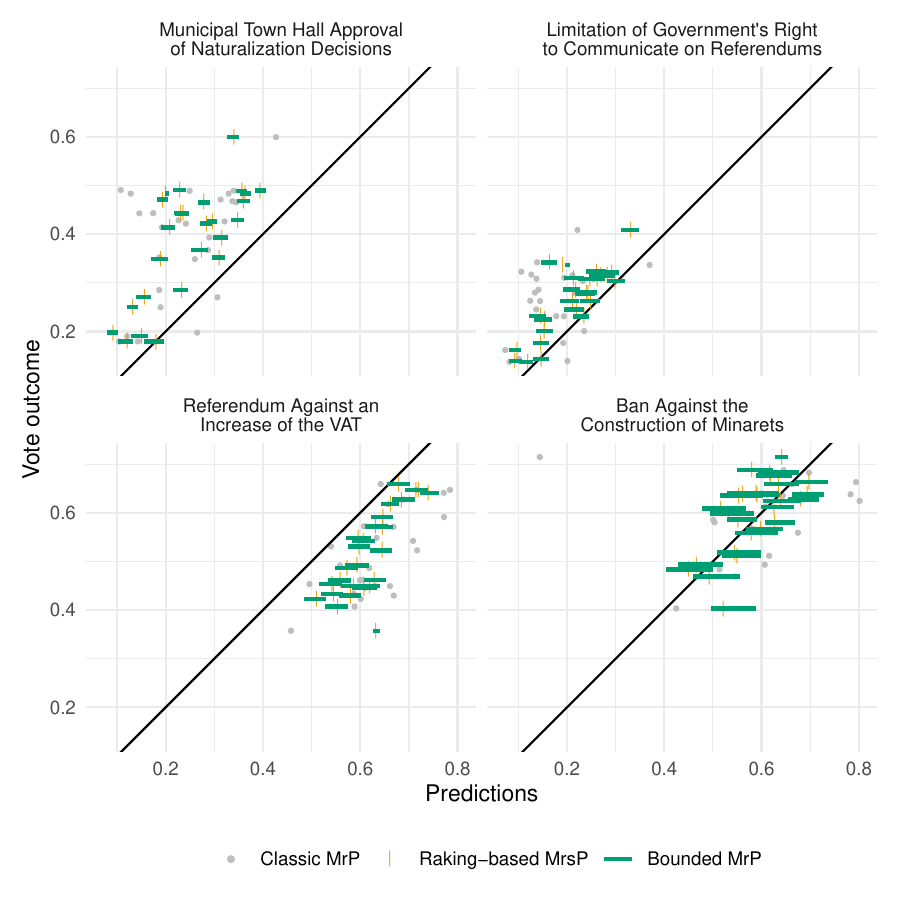}
    \endgraf\vspace{1em}
    \begin{minipage}{\linewidth}\scriptsize
    \textbf{Notes:} Each scatter plot panel corresponds to one referendum and plots MrP/MrsP estimates against the observed cantonal yes-vote share for each of the 26 Swiss cantons. The gray dot marks the classic MrP estimate and the orange tick marks the raking-based MrsP estimate as in the main text; the horizontal green bar spans the MrP bounds computed under the expanded model that allows interactions between party identification and each census-based post-stratification variable. The black diagonal marks perfect prediction.
    \end{minipage}
\end{figure}

Computing the MrP bounds using the same synthetic post-stratification frame, we observe that including interactions widens the bounds considerably: the mean bound width across the four referendums and 26 cantons increases from $0.016$ under the additive model to $0.046$ under the expanded model, almost a threefold increase. Because the interactions between party identification and the census variables are not pinned down by the observed margins, allowing for them enlarges the set of post-stratification distributions---and therefore MrP estimates---consistent with the data. Figure~\ref{fig:app-stress} shows the resulting bounds alongside the classic MrP and raking-based MrsP point estimates from the main analysis.

\section{Additional Analysis}

\subsection{Support for Same-Sex Marriage in the US} \label{sec:app-us}

As a second empirical application, \citet{Leemann.Wasserfallen.2017} estimate support for same-sex marriage across state senate districts (SSDs) in the United States. Their analysis largely replicates that of \citet{Warshaw.Rodden.2012}, who could not include ``age'' as a post-stratification variable because the US census does not publish a cross-tabulation of race, gender, education, and age at the SSD level. \citet{Leemann.Wasserfallen.2017} extend the analysis by combining the SSD-level age marginal with the joint distribution of race, gender, and education. They build an adjusted synthetic post-stratification frame and benchmark their adjusted MrsP estimates against the yes-vote shares from same-sex marriage referendums held in five states (Arizona, California, Michigan, Ohio, and Wisconsin) covering 174 of the 1,942 SSDs. The authors' multilevel probit model includes race, gender, education, and age as individual-level predictors alongside several district-level contextual variables: median income, share of urban residents, share of veterans, share of union members, share of evangelical or Mormon residents, and share of same-sex couples.

\begin{figure}[ht]
    \centering
    \caption{MrP Bounds and MrP Estimates for Same-Sex Marriage Support in 174 US State Senate Districts}
    \label{fig:us-bounds}
    \includegraphics[width=\textwidth]{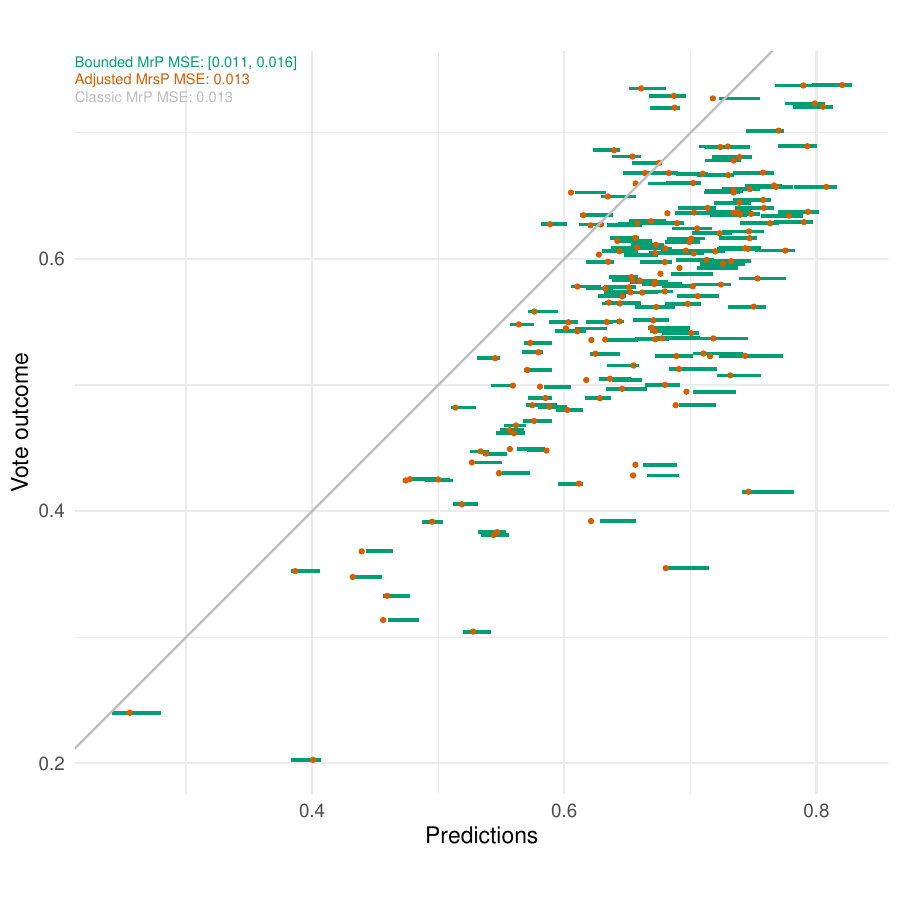}
    \endgraf\vspace{1em}
    \begin{minipage}{\linewidth}\scriptsize
    \textbf{Notes:} Each green bar spans the MrP bounds for one SSD; the orange dot marks the adjusted MrsP estimate. The top-left corner reports, from top to bottom, the achievable MSE range for bounded MrP, the MSE of the adjusted MrsP estimates, and the MSE of the classic MrP estimates.
    \end{minipage}
\end{figure}

We refit the multilevel model on the survey data and then solve the two linear programs separately for each senate district to obtain the MrP bounds. Figure~\ref{fig:us-bounds} plots these MrP bounds against the official referendum yes-vote shares for the 174 benchmark districts, alongside the MrsP point estimates from \citet{Leemann.Wasserfallen.2017}. The achievable MSE range under bounded MrP is $[0.011, 0.016]$, while adjusted MrsP achieves an MSE of $0.013$. Following the authors, we also report the MSE of the classic MrP estimate that omits age from the post-stratification step, as in \citet{Warshaw.Rodden.2012}.

Note that the MrsP MSE of $0.013$ is substantially higher than the value of $0.008$ reported by the authors in Figure~4 of the published paper. The gap traces to a single indexing error in their R code. Their post-stratification frame codes 15 age categories ($20$--$24$ through $90+$, indexed $1$--$15$), while the response model is fit on a survey that also contains a $<20$ category and therefore produces a 16-row table of age random effects (indexed $1$--$16$). The cell-level prediction reads the age random effect with \texttt{ranef(model)\$age.group[cat\_age, 1]}, which for \texttt{cat\_age} $\in \{1,\dots,15\}$ returns rows $1$--$15$ of the 16-row table---i.e., the random effects for the survey's $<20, 20$--$24, \dots, 85$--$89$ groups. Every post-stratification stratum is therefore assigned the age random effect of the next-younger age group, and the $90+$ random effect is never used. Fixing this error eliminates the reported advantage of MrsP over classic MrP.

Two findings emerge from the figure. First, the bounds are narrow and informative about which senate districts are more or less supportive of same-sex marriage. The achievable MSE range for bounded MrP, $[0.011, 0.016]$, contains the MSE of $0.013$ from classic MrP. This contrasts with the Swiss analysis in the main text, where the additional post-stratification variable (party identification) improved the estimates across in three out of four referendums. The reason for the difference is that age is much less predictive of attitudes toward same-sex marriage than party identification is of referendum vote choice in the Swiss data. Second, for some individual senate districts, the MrsP point estimate falls outside the MrP bounds (visible as a gap between the orange dot and the green line segment). This illustrates how the ``adjusted MrsP'' estimates of \citet{Leemann.Wasserfallen.2017} differ from raking-based MrsP estimates: while the latter fall within the MrP bounds by construction, the former do not, because their post-stratification frame is not guaranteed to be consistent with all population margins.

\stopcontents[appendixtoc]

\end{appendices}

\end{document}